\documentclass{ws-ijmpe}
\usepackage{graphicx}
\usepackage{times}
\usepackage{color}
\usepackage[latin1]{inputenc}

\begin{document}

\title{Trionic and quartetting phases in one-dimensional
multicomponent ultracold fermions}
\author{P. Lecheminant}
\address{
Laboratoire de Physique Th\'eorique et Mod\'elisation,
Universit\'e de Cergy-Pontoise, CNRS, 2 Avenue Adolphe Chauvin, 95302
Cergy-Pontoise, France.}
\author{P. Azaria} 
\address{Laboratoire de Physique Th\'eorique de la Matiere
Condens\'ee, Universit\'e Pierre et Marie Curie, CNRS, 4 Place
Jussieu, 75005 Paris, France.}
\author{E. Boulat} 
\address{Laboratoire Mat\'eriaux et Ph\'enom\`enes Quantiques, 
Universit\'e Paris Diderot - Paris
7, CNRS,
75205 Paris Cedex 13, France.}
\author{S. Capponi}
\address{Laboratoire de Physique Th\'eorique, IRSAMC, 
CNRS, Universit\'e Paul Sabatier, 
 31062 Toulouse, France.}
\author{G. Roux} 
\address{Institute for Theoretical Physics C, 
RWTH Aachen University, D-52056 Aachen, Germany.}
\author{S.~R. White} 
\address{Department of Physics and Astronomy, University of California, 
Irvine, CA 92697, USA.}

\maketitle

\begin{history}
\received{(received date)}
\revised{(revised date)}
\end{history}

\begin{abstract}
We investigate the possible formation 
of a molecular condensate, which might be, for instance, the
analogue of the alpha condensate of nuclear
physics, in the context of multicomponent 
cold atoms fermionic systems. 
A simple paradigmatic model of $N$-component fermions with contact
interactions loaded into a one-dimensional optical lattice is studied 
by means of low-energy and numerical approaches. 
For attractive
interaction, a quasi-long-range molecular superfluid phase,
formed from bound-states made of $N$ fermions, 
emerges at low density. 
We show that trionic and quartetting phases, respectively for
$N=3,4$, extend in a large domain
of the phase diagram and are robust against small symmetry-breaking perturbations. 
\end{abstract}

\section{Introduction}

Since the discovery of Bose-Einstein condensation, 
cold atom systems have become
a major field of research for investigating the physics of strong
correlations in a widely tunable range and in unprecedentedly clean
systems.  
In particular, loading cold atomic gases into an optical lattice allows
for the realization of bosonic and fermionic (depending
 on the number of neutrons of the underlying atom) lattice models
and the experimental study of exotic quantum phases in a new
context~\cite{review}.
A prominent example is the observation of the Mott insulator-superfluid
quantum phase transition with cold bosonic atoms in an optical
lattice~\cite{greiner}.
Optical traps offer the opportunity to investigate
the main features of the hyperfine spin ($F$) degeneracy on the properties
of ultracold fermionic quantum gases.
For instance,
three component Fermi gases may be created experimentally
by trapping the lowest three hyperfine states of
$^6$Li ($F=1/2$) atoms in a magnetic field or
by considering $^{40}$K ($F=9/2$) atoms.
In addition, the magnetic field dependence of the 
three scattering lengths of $^6$Li are known experimentally and can
be tuned via Feshbach resonance~\cite{bartenstein} which
opens the experimental realization of a three-component fermionic
lattice model.
In fact, such a degenerate Fermi gas has been created experimentally
very recently \cite{ottenstein}.
The existence of these internal degrees of freedom in fermionic
atoms is expected to give rise to some exotic superfluid phases.
In this respect, a molecular superfluid (MS) phase might be stabilized where
more than two fermions form a bound state.  Such a non-trivial
superfluid behavior has already been found in different contexts.  In
nuclear physics, a four-particle condensate---the $\alpha$ particle---is 
known to be favored over deuteron condensation at low densities~\cite{schuck}. 
This quartet condensation can also
occur in  semiconductors with the formation of
biexcitons~\cite{nozieres}.  A quartetting phase, which stems from the
pairing of Cooper pairs, has also been found in a model of
one-dimensional (1D) Josephson junctions~\cite{doucot} 
and in four-leg Hubbard ladders~\cite{affleckbipairing}. 
A possible experimental observation of quartets
might be found in superconducting
quantum interference devices with (100)/(110) interfaces
of two d-wave superconductors~\cite{schneider}.
The $hc/4e$ periodicity of
the critical current with applied magnetic flux
has been interpreted as the formation of
quartets with total charge $4e$~\cite{aligia}.
More recently, the emergence of
trions (i.e. three-fermion bound states) 
and quartets has been proposed to occur in
the context of ultracold fermionic atoms~\cite{miyake,Rapp,phle,capponipra,Wu2005}.  
A simple paradigmatic model
to investigate the formation of this exotic physics
is the 1D $N$-component fermionic Hubbard model
with attractive (s-wave) contact interaction.
Such a model is defined by the following Hamiltonian:
\begin{equation}
{\cal H}
= -t \sum_{i,\alpha} [c^{\dagger}_{\alpha,i} c_{\alpha,i+1} 
+ {\rm H.c.} ]+\frac{U}{2} \sum_i n_i^2,
\label{hubbardS}
\end{equation}
where $c^{\dagger}_{\alpha,i}$ is the fermion creation operator
corresponding to the $N$ hyperfine states $\alpha = 1,\ldots,N$ and
$n_i = \sum_{\alpha} c^{\dagger}_{\alpha,i} c_{\alpha,i}$ is the
density at site $i$.  
Model~(\ref{hubbardS})  displays an extended
U($N$)$=$U(1)$\times$SU($N$) symmetry 
where the SU($N$) hyperfine spin symmetry is defined by:
$c_{\alpha,i} \rightarrow \sum_{\beta} U_{\alpha \beta}
c_{\beta,i}$ ($U_{\alpha \beta}$ being a SU($N$) matrix).
The MS instability is built from the $N$ fermions: 
$M_i^{\dagger} = c_{1,i}^{\dagger} c_{2,i}^{\dagger}
\cdots c_{N,i}^{\dagger}$ which is a singlet under the SU($N$) symmetry. 
In the following, we will show, by means of a combination of analytical
and numerical results obtained by the density-matrix renormalization group
(DMRG) technique~\cite{dmrg}, that this MS phase emerges
at small enough density $n$ ($n = \sum_i n_i$) for an attractive interaction
($U<0$).
This work is a brief summary of the results obtained in Ref. ~\refcite{capponipra}. It includes also new results concerning
the stability of the MS phase upon switching on 
SU($N$) symmetry-breaking terms for $N=3,4$.

\section{Low-energy approach}

The low-energy effective field theory
corresponding to the SU($N$) model (\ref{hubbardS}) for 
attractive interactions has been discussed 
at length in Refs.~\refcite{phle,capponipra}.
The starting point of this approach is 
the linearization at the two Fermi points
($\pm k_F = \pm \pi n / N a_0$, $a_0$
being the lattice spacing) of the dispersion relation of free $N$-component
fermions~\cite{giamarchi}. 
For incommensurate filling, the resulting low-energy Hamiltonian 
separates into two commuting U(1) density and 
SU($N$) hyperfine spin parts: 
${\cal H} = {\cal
H}_d + {\cal H}_s$.  
This results is the famous ``spin-charge'' separation which
is the hallmark of 1D electronic quantum systems~\cite{giamarchi}.  
Within this low-energy
approach, the U(1) density excitations are critical 
and display metallic properties in the
so-called Luttinger liquid universality class~\cite{giamarchi}. 
In contrast, for 
attractive interaction ($U < 0$),
a spectral gap opens in the hyperfine spin sector.
The dominant instability which governs the physics of 
model (\ref{hubbardS}) is the
one with the slowest decaying correlations at zero temperature.
In Ref.~\refcite{capponipra}, we have found that
the equal-time density correlation $N(x) = \langle n_i n_{i+x} \rangle$
associated to an atomic-density wave (ADW)
and the equal-time MS correlations $M(x) = \langle
M_i M_{i+x}^{\dagger}\rangle$ display  
the following power-law decay at long distance:
\begin{eqnarray}
N(x)& \sim& \cos(2 k_F x) \; x^{-2 K/N} \label{dens.eq}\\
M(x) &\sim&  x^{-N/(2 K)} \quad \text{for }N\text{ even}\label{Neven.eq}\\
M(x) &\sim& \sin(k_F x) \; x^{-(K+N^2/K)/(2N)}\quad
\text{for }N\text{ odd,}\label{Nodd.eq}
\end{eqnarray}
where $K$ is the Luttinger parameter which depends
on the interaction $U$ and density $n$.
It is a non-universal parameter which controls
the power-law decay of correlation functions
and is central to the Luttinger-liquid paradigm~\cite{giamarchi}.
A perturbative estimate of $K$ can be determined:
$K = \left[1 + U (N -1)/(\pi
v_F)\right]^{-1/2}$ ($v_F = 2t a_0 \sin (\pi n/N)$
being the Fermi velocity). 
We thus observe that ADW and MS instabilities 
compete and the key point of the analysis is the one
which dominates. 
In particular, we deduce from Eqs. (\ref{dens.eq}, \ref{Neven.eq},
\ref{Nodd.eq}) that a
dominant MS instability requires   $K > N/2$ 
($K > N/\sqrt{3}$) for $N$ even (odd respectively). 
The existence and stability of this MS phase stem from the 
knowledge of the full
non-perturbative behavior of the Luttinger parameter $K$ as a function
of the density $n$ and the interaction $U$. 
This parameter can be numerically determined 
in the simplest cases
$N=3,4$ by computing dominant correlations with the 
DMRG technique.
\begin{figure}[t]
\includegraphics[width=\columnwidth,clip]{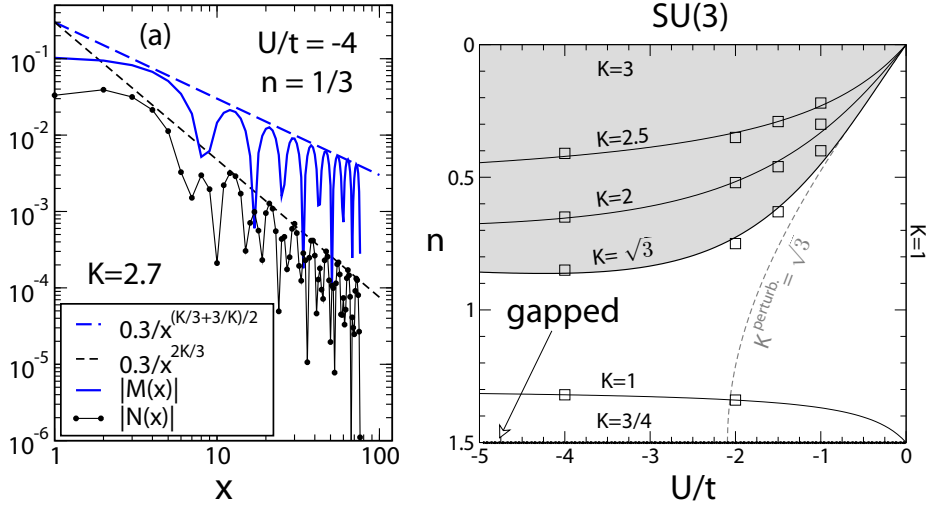}
\caption{(Color online) SU(3) model: trions and density correlations
vs distance obtained by DMRG with $L=153$, $n=1/3$ and $U/t=-4$. (a) dominant
trions over ADW correlations can both be fitted with $K=2.7$. 
(b) phase diagram showing the Luttinger parameter
$K$ vs filling $n$ and interaction $U$. The grey area is the
trionic phase. Lines are guide for the eyes.
}
\label{CorrSU3.fig}
\end{figure}
\begin{figure}[t]
\includegraphics[width=\columnwidth,clip]{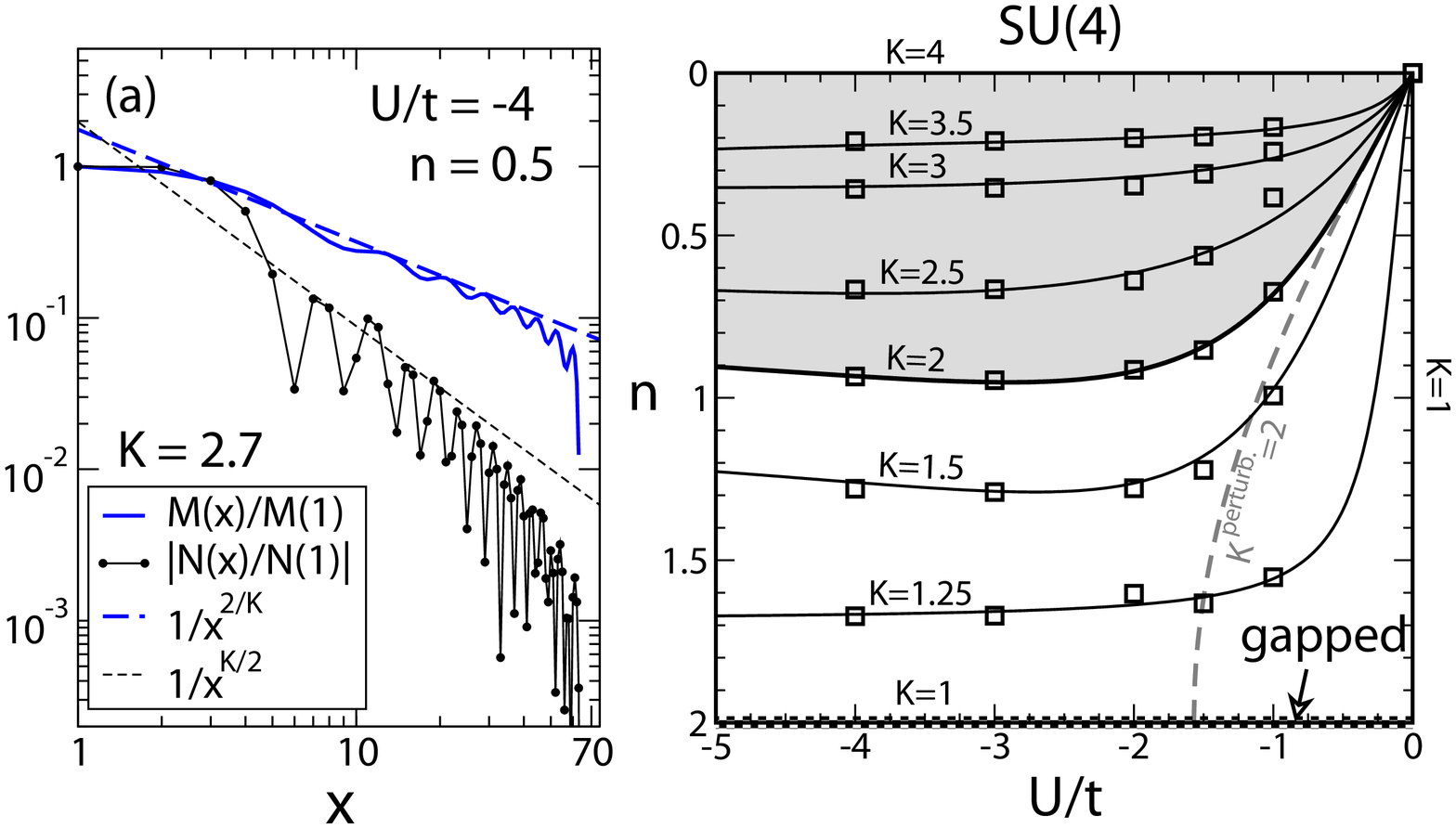}
\caption{(Color online) SU(4) model: (a) quartet and density
correlations vs distance obtained by DMRG with $L=128$ and $U/t=-4$ at
filling $n = 0.25$. 
(b) phase diagram showing the Luttinger parameter
$K$ vs filling $n$ and interaction $U$. The grey area is the
quartetting phase. Lines are guide for the eyes.
 }
\label{CorrSU4.fig}
\end{figure}

\section{DMRG results}

We have performed extensive DMRG
calculations with open-boundary conditions
for both the $N=3$ and $N=4$ cases and for a wide range
of densities $n$ and interactions $U$. 
We see in Fig.~\ref{CorrSU3.fig}(a) and
Fig.~\ref{CorrSU4.fig}(a)  that the density and MS
correlations  $N(x)$ and $M(x)$
display a power-law behavior for
$N=3$ and $N=4$ respectively at  typical values of  $n$
and $U= -4t$.
The phase
diagrams for SU(3) and SU(4) models are presented in 
Figs.~\ref{CorrSU3.fig}(b), \ref{CorrSU4.fig}(b)
which give a map of $K$ vs interaction and density. 
The values of $K$ were obtained from the power-law behavior of the molecular
correlation $M(x)$ using Eqs.~(\ref{Neven.eq}, \ref{Nodd.eq}).  
We find that 
trions and quartet superfluid phases emerge 
in a wide portion of the phase diagrams (grey area) at low density 
separated from a ADW phase by a cross-over line $n_c(U)$.  
Interestingly enough,
the MS phase extends to small values of $U$ at sufficiently
small densities. 
In absence of the optical lattice, i.e. $n=0$, 
we recover the known results that trionic and quartetting
phases are stabilized respectively for $N=3$ and $N=4$.
Indeed, 
the continuum SU($N$) fermionic model with a delta interaction 
is integrable by means of the Bethe ansatz 
and a phase with the formation of bound-state of $N$ fermions is found
for an attractive interaction~\cite{betheansatz}.

\begin{figure}[t]
\includegraphics[width=\columnwidth,clip]{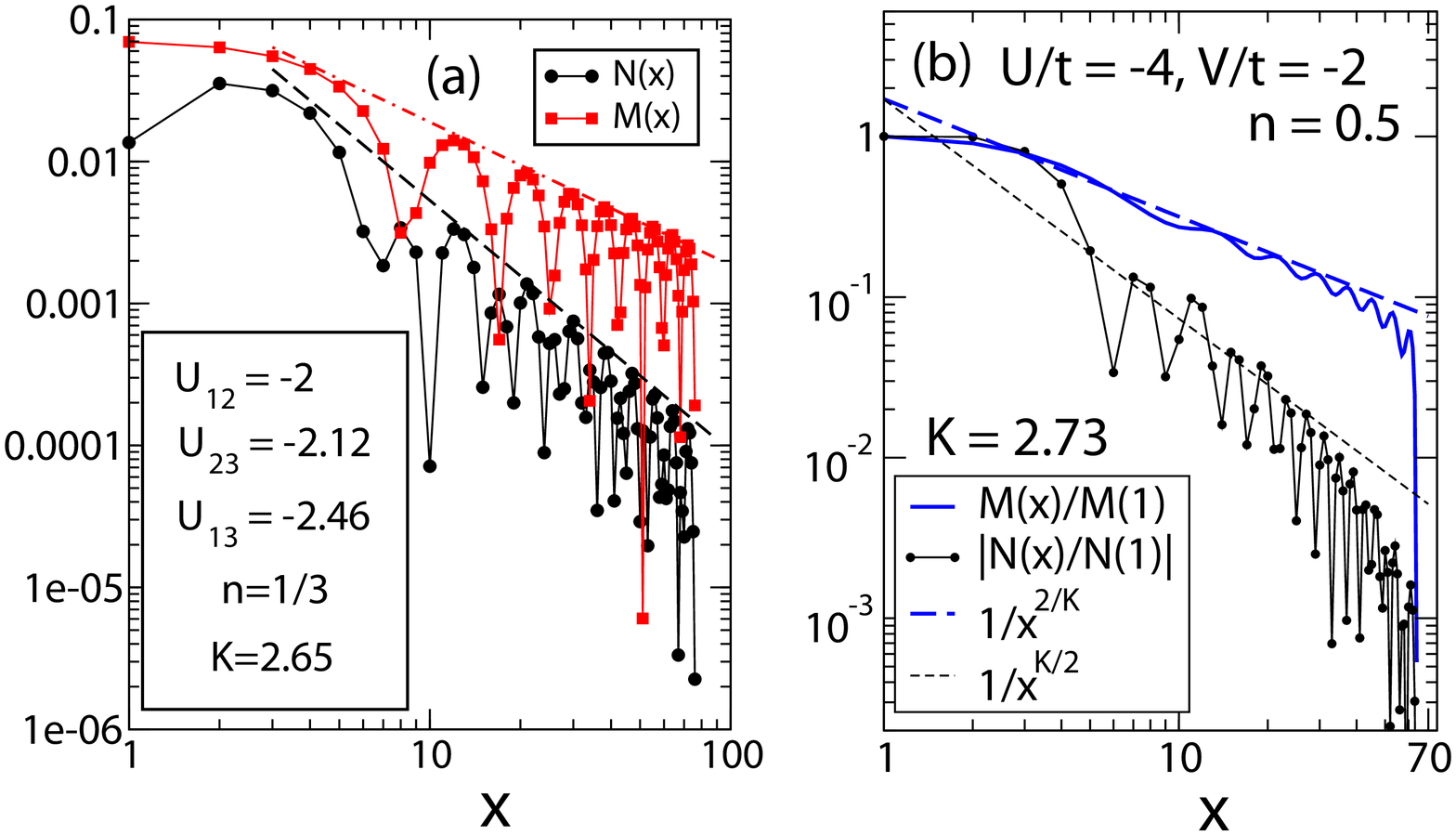}
\caption{
(a): Trions and density correlations for the anisotropic
three-components fermionic model (\ref{hubbard3}) at density $n=1/3$.
The value of the Luttinger parameter is $K=2.65 > \sqrt{3}$
which signals the onset of the trionic phase.
(b):
Quartet and density correlations for the SO(5) model with $U/t=-4$ and
$V/t=-2$ at the density n=0.5.
The value of the Luttinger parameter is $K=2.7 > 2$ which shows
that the quartetting phase survives the breaking of the SU(4)
symmetry.
}
\label{CorrWithBCS_SO5.fig}
\end{figure}


\section{Effect of symmetry breaking perturbations}
At this point, the natural question is whether the MS 
phases survives to the breaking of the artificial SU($N$) 
symmetry of model (\ref{hubbardS}).  
In the $N=3$ case, the basic assumption of the SU(3) symmetry
is that the three scattering lengths of the underlying 
cold atoms problem are the same. 
For the $^6$Li atoms, these scattering lengths are different
and can be varied with an external magnetic field~\cite{bartenstein}.
The minimal model to investigate the low-energy properties of 
$^6$Li fermionic atoms in 1D reads as follows:
\begin{equation}
{\cal H}
= -t \sum_{i,\alpha} [c^{\dagger}_{\alpha,i} c_{\alpha,i+1}
+ {\rm H.c.} ]+ U_{12} \sum_i n_{i,1} n_{i,2}  
+ U_{13} \sum_i n_{i,1} n_{i,3} + 
U_{23} \sum_i n_{i,2} n_{i,3},
\label{hubbard3}
\end{equation}
where $n_{i,\alpha} = c^{\dagger}_{\alpha,i} c_{\alpha,i}$ is
the density of the fermion on site $i$ with
hyperfine spin component $\alpha =1,2,3$.
We observe that the SU(3) symmetry of model (\ref{hubbardS}) 
is broken down to $U(1)^{3}$.
In Fig. 3 (a), we show for an anisotropic situation close to the
SU(3) point, which is relevant to 
the $^6$Li problem, that the trionic phase survives the breaking of the
artificial SU(3) symmetry.
The investigation of the full phase diagram of model (\ref{hubbard3}) 
at zero temperature will be discussed in a forthcoming paper.
For the $N=4$ case, 
we consider the model for spin-3/2 (i.e. four-component) fermionic cold atoms
with s-wave contact interaction derived in  Ref.~\refcite{ho}.
The interacting part depends on $U_J$ parameters corresponding
to the total spin $J$ of two spin-3/2 atoms which
takes only even integers value 
due to Pauli's principle (the hyperfine spin part of the wave
function should be
antisymmetric): $J=0,2$.
The resulting model reads as follows:
\begin{equation}
{\cal H}
= -t \sum_{i,\alpha} [c^{\dagger}_{\alpha,i} c_{\alpha,i+1}
+ {\rm H.c.} ]+\frac{U}{2} \sum_i n_i^2
+ V  \sum_{i}
P_{00,i}^{\dagger} P_{00,i},
\label{hubbard3demi}
\end{equation}
where $c^{\dagger}_{\alpha,i}$ is the fermion creation operator
corresponding to the spin-3/2 states $\alpha = \pm 3/2, \pm 1/2$ and
$P_{00,i}^{\dagger} = c^{\dagger}_{3/2,i}c^{\dagger}_{-3/2,i} -
c^{\dagger}_{1/2,i}c^{\dagger}_{-1/2,i}$
is the singlet-pairing operator with $V = U_0 - U_2$.
As shown
in Ref.~\refcite{zhang}, the presence of this 
BCS singlet-pairing term reduces the SU(4) symmetry down 
to SO(5) when $V \ne 0$.  We  show typical data
for $U/t = -4$ and $V/t =-2$ at the density $n=1/2$ 
in Fig.~\ref{CorrWithBCS_SO5.fig} (b).
Quartet correlations are (quasi) 
long ranged and dominate over ADW ones. 
For large negative $V$, 
a BCS pairing phase emerges~\cite{Capponi2007} but the main point here
is to show that  the  quartetting phase is not an artifact
of the SU(4) symmetry and 
does exist in more realistic models at low density.

\section{Concluding remarks}

We have shown that a quasi-long-range
MS phase can emerge in 1D multicomponent fermionic
model with attractive contact
interaction at sufficiently low density. This phase, characterized 
by the formation of a bound-state made of $N$ fermions,
is not an artifact of the
artificial SU($N$) symmetry of model (\ref{hubbardS}). In this
respect,  the trionic and
quartetting phases in the simplest $N=3,4$ cases might be explored
experimentally in the context of spinor ultracold fermionic atoms.  As
a first step, we have assumed here a homogeneous optical lattice and
have neglected the parabolic confining potential
of the atomic trap.  We expect that this potential will not affect
the main qualitative properties of this MS phase at low density. The effect
of a harmonic trap could be investigated by means of DMRG calculations for
quantitative comparisons as it has been done for 1D two-component 
fermionic systems~\cite{trap}. 
Such a study is currently under progress in the $N=3,4$ cases.
We hope that future experiments in ultracold spinor
fermionic atoms will reveal the existence of these trionic and quartetting
phases.

\section*{Acknowledgements}
We would like to thank  J. Dukelsky and P.~Schuck,
for useful discussions
and their interest. 
SC and GR thank IDRIS
(Orsay, France) and CALMIP (Toulouse, France) for use of
supercomputers. SRW acknowledges the support of the NSF under
grant DMR-0605444.

\end{document}